\documentclass[useAMS,usenatbib]{mn2e}
\title[Quantification of stochastic fragmentation of self-gravitating discs]{Quantification of stochastic fragmentation of self-gravitating discs}
\author[Young et al.]{
M.D.~Young$^1$, 
and
C.J. Clarke$^1$ \\
$^1$Institute of Astronomy, University of Cambridge, Madingley Road, Cambridge,
CB3 0HA, United Kingdom\\
}
\usepackage[pdftex]{graphicx}
\usepackage{natbib}
\usepackage{url}
\usepackage{amsmath}
\usepackage{amssymb}
\usepackage{caption}

\newcommand{\abs}[1]{\left| #1 \right|} 
 
\let\baraccent=\= 
\renewcommand{\=}[1]{\stackrel{#1}{=}} 

\begin{document}
\date{Written June 2015}
\maketitle
\begin{abstract}

Using 2D smoothed particle hydrodynamics, we investigate the distribution of
wait times between strong shocks in a turbulent, self-gravitating accretion
disc.  We show the resulting distributions do not depend strongly on the
cooling time or resolution of the disc and that they are consistent with
the predictions of earlier work \citep{MYfrag,Cossins1,Cossins2}.  We use
the distribution of wait times between shocks to estimate the likelihood of
stochastic fragmentation by gradual contraction of shear-resistant clumps on
the cooling time scale.  We conclude that the stochastic fragmentation
mechanism \citep{PKstochastic} cannot change the radius at which
fragmentation is possible by more than $\sim 20\%$, restricting direct
gravitational collapse as a mechanism for giant planet formation to the outer
regions of protoplanetary discs.

\end{abstract}

\section{Introduction}
\label{sec:intro}

It is widely believed that planets can form via fragmentation of disc gas only in
the outermost regions of protoplanetary discs (i.e. $>  50$ A.U.; 
\cite{Rafikov2005,CathieChemFrag,Rice09}) . This mechanism is thus favoured 
for explaining planets in wide orbits as detected via direct imaging 
\citep{HR8799,HR8799_2} for which the time to form by the alternative core 
accretion model is uncomfortably long.  Core accretion (i.e. the assembly of 
a rock core and subsequent accretion of disc gas) is therefore favoured for 
the formation of the vast majority of gas giant planets detected to date.

The argument against forming planets by fragmentation in the inner
disc hinges on the fact that self-gravitating discs are very optically
thick at smaller radii which renders $\beta$ (the ratio of cooling
time to dynamical time) very large. A large number of numerical
experiments have shown that whether a self-gravitating disc fragments
or whether it is maintained in a self-regulated (non-fragmenting) state
depends on the value of $\beta$. We will not here repeat the
considerable debate about the exact value of the critical $\beta$
value for fragmentation, nor its dependence on numerical technique and
resolution
\citep{BetaCooling,Rice05,Gammie01,MeruBate1,MeruBate2,honHRes,MYfrag,RiceCool,RiceCool2}. 
Suffice it to say that  the range of values of
$\beta_{crit}$ claimed in the literature ($3-30$), corresponds to a
rather narrow range of minimum radii for fragmentation because the
radial dependence of $\beta$ in the outer regions of a steady state self-gravitating disc
is steep ($\beta \propto r^{-4.5}$) \citep{Rafikov2005,Rafikov2009,CathieChemFrag}.

All of the above numerical experiments refer to {\it prompt} fragmentation
(i.e. fragmentation that occurs within a cooling time of the disc being
set up in a self-gravitating state). \cite{PKstochastic} however, reported a
distinct fragmentation mode in long timescale integrations using the
grid code FARGO 2D. In this experiment the disc apparently settled into
the self-regulated state and then eventually formed a fragment after a 
large number of cooling times. He dubbed this alternative mode of 
fragmentation ``stochastic fragmentation''.

The prospect of eventual fragmentation even at high $\beta$ values
could potentially over-turn the conclusions drawn above about the impossibility 
of forming planets by gravitational fragmentation at small radii.  Given that 
the self-gravitating lifetime of a disc is $> 10^4$ dynamical times at the 
radius of Jupiter, it follows that even an extremely small (but non-zero) 
probability of fragmentation per dynamical time could lead to planet 
formation by this route.

The possibility of stochastic fragmentation in self-gravitating media
has also been discussed by \cite{Hopkins}.
In their picture, stochastic fragmentation may result from the collapse 
of extremely rare non-linear density fluctuations on a scale much smaller 
than the Jeans length; their study quantifies the probability
of this outcome using the statistics of isothermal turbulence.
Although those studies that have attempted to quantify the power-spectrum of
the turbulence have found little power at sub-Jeans scales
\citep{Boley07,Cossins1}, we cannot rule out the possibility that such a mechanism 
is operative in self-gravitating discs (indeed no simulations performed to date
have the extremely sub-Jeans length resolution required for this
analysis).  However, this is definitely {\it not} the mechanism that is
operating in the simulations of \cite{PKstochastic}.  Here collapse
is initiated on the Jeans scale (as expected) and the proto-clump
contracts on the (long) cooling time of the simulation.
With such a very slow contraction, the expected outcome is that
proto-clumps are disrupted by collisions with the spiral arms. The
``stochastic'' (and unexpected) outcome of the Paardekooper simulation
was that clumps can occasionally avoid such collisions over a sufficiently 
long time window that they manage to collapse.

We have argued above that it is essential to quantify the probability
of stochastic fragmentation before ruling out fragmentation of the inner disc. 
Such quantification is however extremely difficult through direct simulation. 
By definition, stochastic fragmentation simulations must employ large $\beta$ 
values in order to avoid the well studied prompt fragmentation regime. Spiral
structure is weak at large $\beta$ \citep{Cossins1} and so
high resolution is required in order that weak spiral heating
is not swamped by artificial viscosity \citep{honHRes}. The nature of stochastic 
fragmentation also implies that long integration times (many cooling times) are required. 
The combination of high resolution and long integration times implies that direct
simulation is not likely to be a feasible strategy in the foreseeable future.

Here we adopt a different approach. We argued in \cite{MYfrag} that the 
$\beta$ threshold for prompt fragmentation can be simply understood in terms 
of the {\it mean} spiral structure in self-gravitating discs.  \cite{Cossins1}
presented a detailed characterisation (wavenumbers, pattern speed, pitch angles) of
the spiral modes present in such discs which allows the calculation
of the {\it mean time} between spiral arm passages. \cite{MYfrag}
interpreted the fragmentation boundary as corresponding to the criterion that a 
proto-fragment can cool on a timescale less than this mean inter-spiral time; 
if this condition is not fulfilled then proto-fragments are disrupted by 
spiral arm passage before they collapse to the small scales where they are 
immune to such disruption.

If this were the full story then there would be no possibility of stochastic 
fragmentation at larger $\beta$. However a casual inspection of simulations of 
self-gravitating discs is enough to show that individual spiral features form, 
dissolve and re-form on a timescale that is close to dynamical. Although the
\cite{Cossins1} description is a reasonable representation of the time-averaged 
behaviour, it does not address the possible {\it variability} in the 
inter-spiral time. In the fluctuating density field encountered in the 
simulations it may be possible for individual proto-fragments to avoid spiral 
shocks over significantly longer periods.

In this paper we therefore analyse the intermittent nature of
spiral structure in self-gravitating discs, recording the
distribution of inter-spiral times for fluid elements in the disc.
We expect from \cite{Cossins1} that the {\it geometry} (though not the amplitude)
of spiral features is not a strong function of either $\beta$ or resolution
but test this hypothesis with a range of simulations. We emphasise that we are
{\it not} seeking to model fragmentation itself (and choose simulation
parameters with sufficiently high $\beta$ that prompt fragmentation is
avoided) but instead use the Lagrangian nature of SPH to analyse particle
histories and infer the distribution of wait times between spiral arm
encounters. In going on to draw conclusions about the survival prospects of any
proto-fragments that might eventually form in the disc, we will be assuming
that such fragments would be sampling the same fluctuating density field that
we analyse here. 

\section{Disc model}
\label{sec:model}

We construct a disc with a power law surface density profile, assume a locally
isothermal Gaussian in the vertical direction with scale height $H$ and set the
initial temperature such that $Q=Q_0$ at all radii.  To ensure the simulation
is scale free and the resolution $h/H$ and aspect ratio $H/R$ are independent
of radius, we choose a surface density power law index of $-2$ \citep{MYfrag}.  
This choice gives,
\begin{eqnarray}
  \Sigma & = & \frac{M_D}{2\pi\log(\xi) R^2} \\
  c_s & = & \sqrt{\frac{GM}{R}} \frac{qQ_0}{2\log(\xi)} 
  \label{eq:initCond}
\end{eqnarray}
where $\Sigma$ is the surface density, $M$ is the star's mass, $M_D$ the disc 
mass, $q=M_D/M$, $c_s$ is the sound speed of the gas, and $Q_0$ is the initial
value of $Q$. $\xi=R_o/R_i$ where $R_o$ and $R_i$ are the outer and inner
radii of the disc respectively.  The disc's aspect ratio is,
\begin{equation}
  \frac{H}{R} = \frac{qQ_0}{2\log(\xi)}
  \label{eq:HonR}
\end{equation}
and the ratio of SPH smoothing length to disc scale height, $h/H$ is,
\begin{equation}
  \frac{h}{H} = \left( \frac{8\pi\log(\xi)^3}{Nq^2Q_0^2} \right)^{1/2}
  \label{eq:honH}
\end{equation}

Our simulations were performed using the same modified version of GADGET2
\citep{Gadget2Code} used by \cite{MYfrag}.  The code modifications include 
artificial conductivity \citep{RiemannSPH}, $\beta$ cooling (i.e. loss of
internal energy on a time scale that is a fixed multiple ($\beta$) of the
local dynamical time), particle accretion 
and the correct treatment of softening with variable smoothing lengths
\citep{gravSoftTerms} and the artificial viscosity 
method of \cite{Cullen}.  We set the artificial viscosity (and
conductivity) to the values that produce the best results in test problems
where the correct result is known (e.g., shock tube test, Sedov blast wave,
Kelvin-Helmholtz instability).  That is, we set $\alpha_{cond}=1.0$, 
$\alpha_{max} = 5.0$,$\alpha_{min}=0.0$ and $l=0.05$ \citep{Cullen}.  
Note that the high value of $\alpha_{max}$ translates in practice to an
average per-particle value of $\alpha_{SPH} \sim 0.1$ away from shocks.
In order to include the effects of the vertical distribution of mass in our
calculation of gravity, we softened all gravitational interactions on the
scale $H$ using the same method as \citep{MYfrag}.  

All our simulations were run in 2D, with $\xi=5$ and $q=0.2$ to minimise
computational expense \citep{MYfrag}.  Each simulation was run for at least
$10$ cooling times at the outer edge, ensuring that the disc reached a
``settled'' state.  We limited our analysis to the final 
$500t_{dyn}$ of each simulation ($200t_{dyn}$ for the highest resolution run),
where the disc was in the $Q\sim1$ gravo-turbulent state.

\section{Results}
\label{sec:shocktime}

\cite{PKstochastic} and \cite{MYfrag} interpreted stochastic fragmentation as arising 
from unstable over-densities that survive for long enough to become bound and resist
disruption.  That is, the gravo-turbulent, $Q \sim 1$ state is constantly
producing gravitationally unstable over-densities.  The majority of these
over-densities are disrupted before they can become bound enough to survive
disruption by the environment.  If an over-density is ``lucky'' and
can survive for long enough without being disrupted, it will form a fragment.

\cite{PKstochastic} found that over-densities consistent with this explanation 
could be found at all points in high resolution simulations which attain 
$Q \sim 1$ but do not promptly fragment and that the number of 
these ``potential fragments'' declined as $1/\beta$.  This heuristic picture was 
further developed by
\cite{MYfrag}, who proposed that the survival of clumps depended on their ability 
to survive encounters with spiral shocks.  That is, for fragmentation to occur, an
over-density must be able to collapse sufficiently before encountering a spiral
shock.  On average, 
the time between successive spiral shocks can be shown to be,
\begin{equation}
  t_{spi} = \frac{2\pi}{m\xi} t_{dyn}
  \label{eq:tspi}
\end{equation}
where $t_{dyn}=\Omega^{-1}$, $m$ is the azimuthal wavenumber of the spiral pattern,
$\Omega_p$ is its pattern speed and $\xi = \abs{\Omega_p-\Omega}/\Omega$
\citep{MYfrag}.   Realistic disc assumptions yield $t_{spi} = 9-13 \Omega^{-1}$, although Equation
\ref{eq:tspi} is only likely to be accurate to about a factor $2$.

To quantify the likelihood of stochastic fragmentation, we measured the
distribution of times between successive spiral shock wave encounters for many
particles within our quasi-stable simulations.  Because reliably detecting
shock fronts is numerically challenging \citep{Cullen,AVMM}, we only consider
the relatively strong shocks in the disc.  We tracked individual particles
over roughly $45$, $90$, $180$ or $360$ dynamical times and marked all periods 
of increase in the particle's entropy and surface density (some of which 
spanned multiple snapshots) as potential encounters with shocks.  We deemed a shock to have
been encountered whenever a particle's entropy and surface density increased by more 
than the median increase across all potential shocks \footnote{The maximum
strength of shocks in a gravito-turbulent disc decreases with $\beta$
\citep{Cossins1}.  Because of this, it is necessary to define what constitutes
a ``strong shock'' relative to the distribution of shocks in a given
simulation, as we have done here.}.  We limited our analysis to particles within
$R=2-3 R_{i}$ to avoid corruption by edge effects.

\begin{figure}
  \begin{center}
    \includegraphics[width=0.5\textwidth]{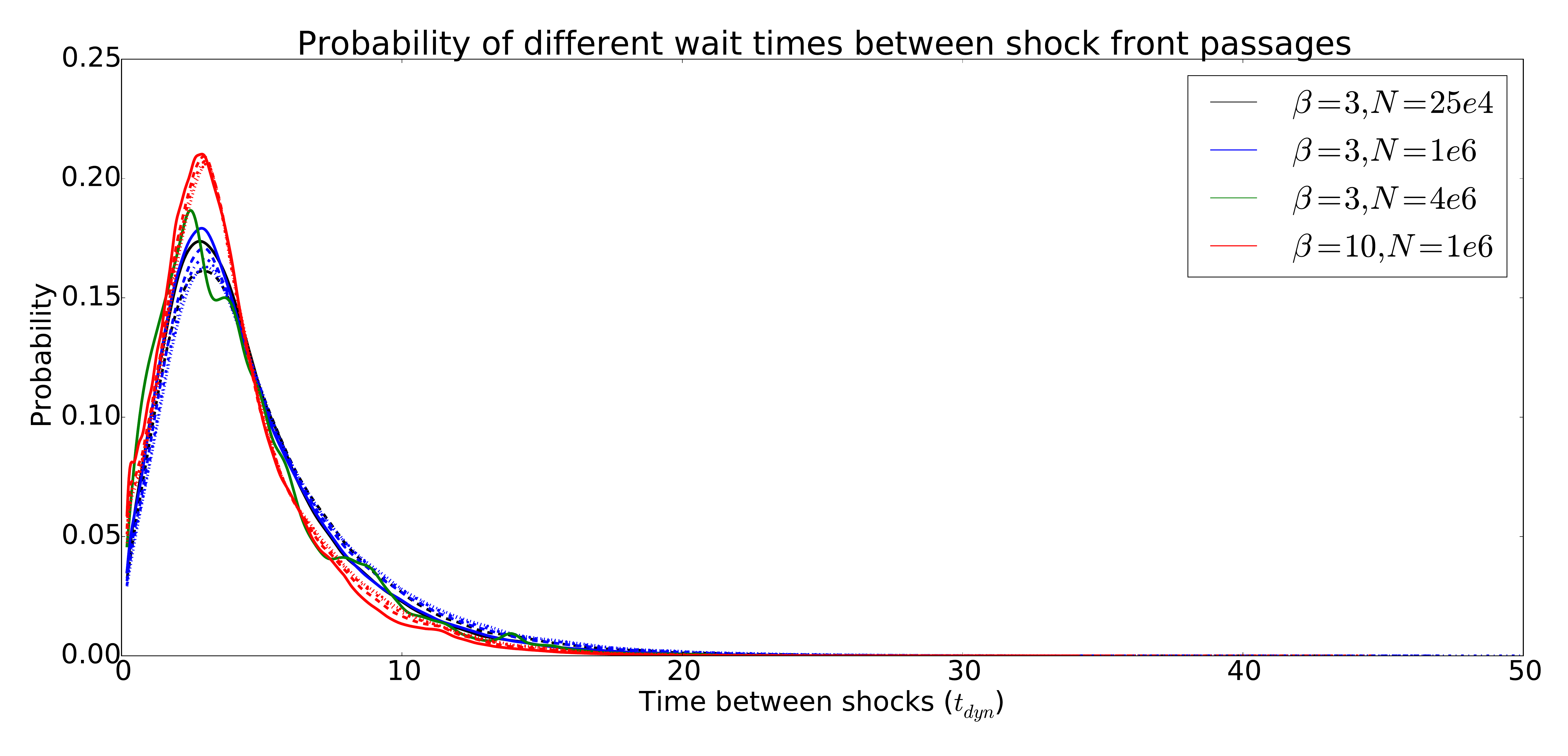}
  \end{center}
  \caption{Distributions of wait times between shocks for simulations at
  different resolutions and $\beta$, inferred by tracking particles for
  different lengths of time.  Particles were tracked across $128$
  (solid lines), $256$ (dashed lines), $512$ (dot-dashed lines) and $1024$
  (dotted lines) dynamical times at the disc's inner edge and all increases of
  entropy were recorded.  A shock was deemed to have been
  encountered whenever the entropy and surface density increased by more than the
  median increase amongst all potential shocks (see text for details).}
  \label{fig:tshocks}
\end{figure}

Figure \ref{fig:tshocks} shows the resulting distributions of wait time
between shocks for simulations at a range of resolutions, values of
$\beta$ and particle tracking times.  There is no significant difference in
the average wait times between simulations
at different resolutions or with different cooling rates.  Furthermore, the distributions
are converged with respect to the length of time a particle is tracked
for, indicating that neither increased resolution nor longer simulation runs would
change the result.

\begin{figure}
  \begin{center}
    \includegraphics[width=0.5\textwidth]{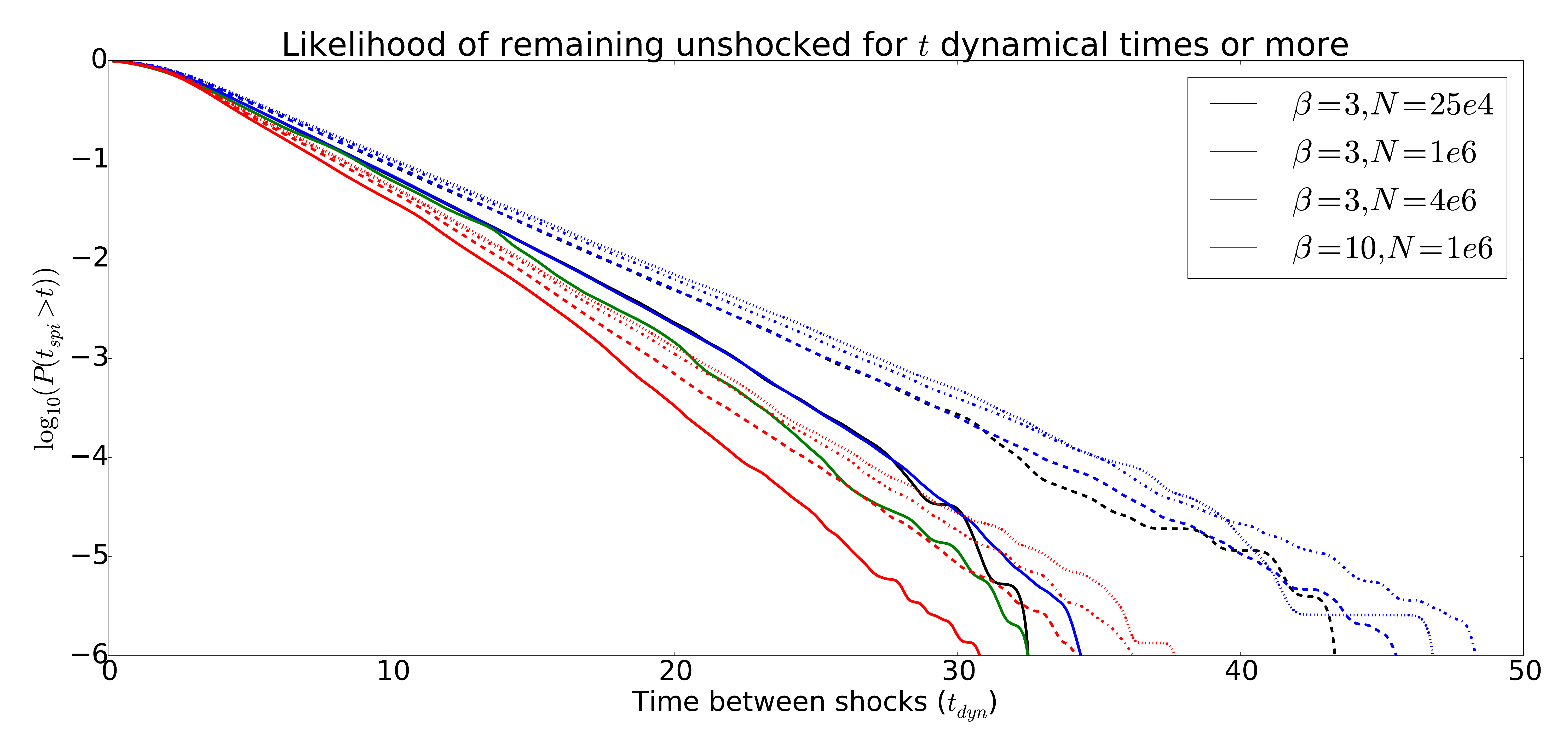}
  \end{center}
  \caption{$\log_{10}$ of the probability that the time between shocks will be
  at least the number of dynamical times shown on the x-axis.  Particles were tracked across $128$
  (solid lines), $256$ (dashed lines), $512$ (dot-dashed lines) and $1024$
  (dotted lines) dynamical times at disc's the inner edge.  See text and
  Figure \ref{fig:tshocks} for further details of how the time between shocks
  was calculated.}
  \label{fig:logp}
\end{figure}

Most importantly, all our simulations show an exponential decay in the
likelihood of a patch of disc remaining unshocked, with long periods of time
(100s of dynamical times or more) having effectively zero probability in all
cases. This is shown more clearly in Figure \ref{fig:logp}, which shows the $\log_{10}$ cumulative
probability of surviving for longer than a certain number of dynamical times
without encountering a shock.  The slope change at the extreme tail of each
distribution is caused by low number statistics resulting from our finite
sample of particles and finite tracking time.  However, it is clear that
the probability of a clump surviving for longer than $\sim 10 t_{dyn}$ without
encountering a strong shock decreases exponentially.

We now relate this wait time distribution to the fragmentation probability per
unit time at radius $R$ ($\dot P_{\rm frag}(R)$).  This involves assumptions
about the numbers of ``potential fragments'' produced per unit time and also a
criterion for the collapse of such potential fragments.  In regard to the
former, \cite{PKstochastic} found that the frequency with which such clumps
form scales as $1/\beta$.  Regarding the collapse criterion, we first assume
that a fragment can collapse only if it does not encounter a shock within a
cooling timescale of formation.  Since such clumps collapse quasistatically on
the cooling time, this criterion is equivalent to requiring that the fragment
is able to change its mean density by order unity before encountering a spiral
shock.  That is, it is well within its Hill sphere at the point that it
receives thermal energy input from the shock interaction (\cite{MYfrag}: we
will experiment with a more permissive requirement for collapse below).  In
the outer parts of the disc, the cooling time scales as $\beta \sim R^{-9/2}$
\citep{Rafikov2005,CathieChemFrag,Cossins2,PKstochastic} and we thus deduce
that  
\begin{equation}
  \dot P_{\rm frag}(R) \propto \left( \frac{1}{\beta} \right) \exp(-A\beta) 
  \propto R^{9/2} \exp(-A \times 20 R_{50}^{-9/2})
  \label{eq:pfrag}
\end{equation}
The exponential term reflects the tail of the cumulative wait time
distribution shown in Figure \ref{fig:logp} and we use Figure \ref{fig:logp}
to derive a value of $A \approx 0.2$.  We have furthermore normalised the
radial dependence of $\beta$ in the outer regions of a self-gravitating disc
such that $\beta=20$ at $R_{50}= R/(50 {\rm A.U.})=1$.  If a disc remains 
self-gravitating for a time $t_{disc}$, Poisson statistics implies that the
probability of its fragmentation over this timescale is,
\begin{equation}
  P_{frag}(R) = 1-\exp(-\dot P_{frag}(R)t_{disc})
  \label{eq:ptot}
\end{equation}
\cite{PKstochastic} found that when $\beta=9$ the probability of stochastic
fragmentation per dynamical time was roughly $1/1700$, which can be used to
normalise Equation \ref{eq:pfrag}.  

\begin{figure}
  \begin{center}
    \includegraphics[width=0.5\textwidth]{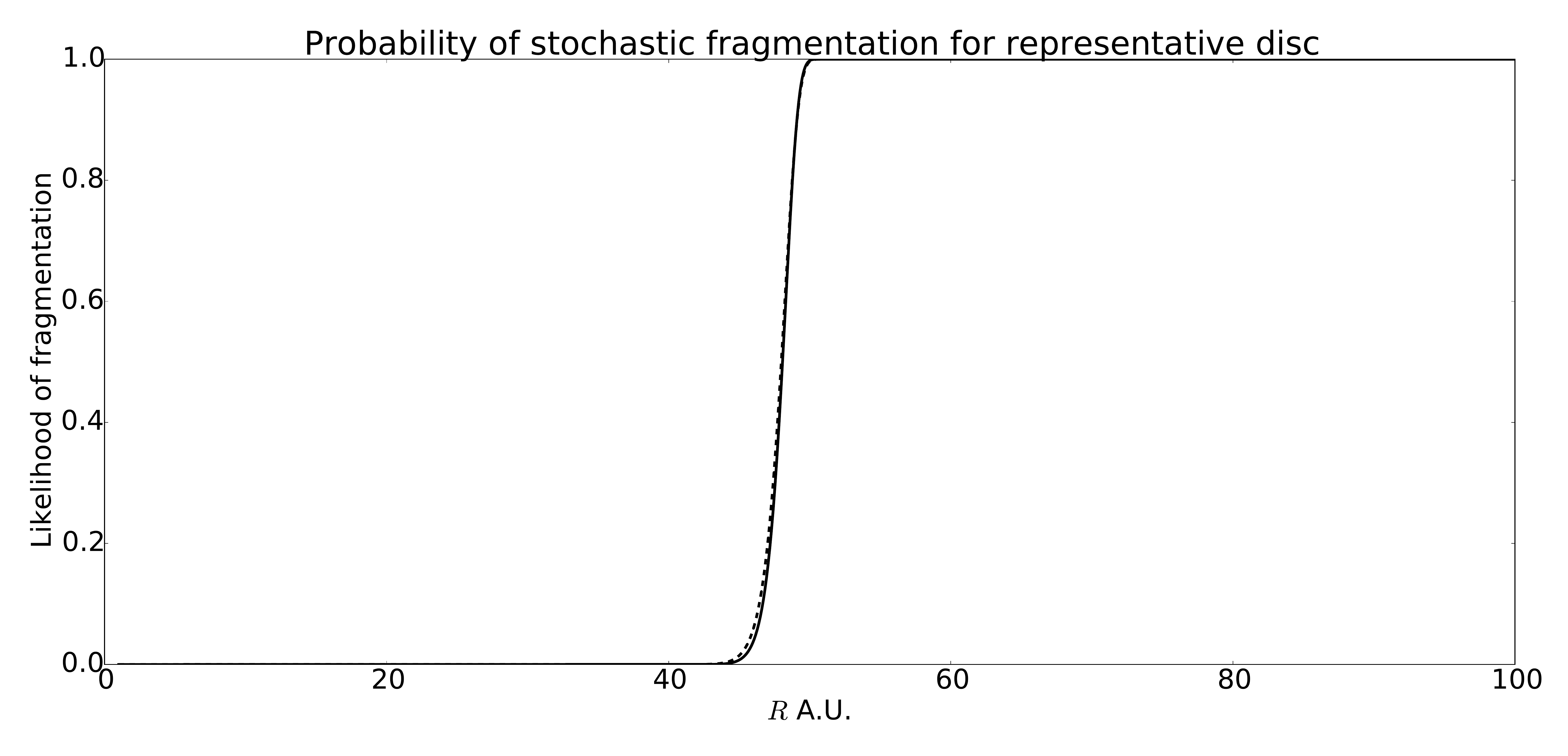}
  \end{center}
  \caption{The probability of stochastic fragmentation occurring during the
  self-gravitating lifetime of a disc at different radii calculated from 
  Equation \ref{eq:ptot} for a disc around a solar mass star
  with a self-gravitating phase lasting $100,000$ years.  The solid and dashed
  lines show the probability of fragmentation assuming, respectively, that 
  fragmentation occurs when a clump survives for more than $t_{cool}$ and that
  fragmentation occurs when a clump exceeds a certain critical density.}
  \label{fig:scale}
\end{figure}

Figure \ref{fig:scale} shows the 
probability of fragmentation as a function of radius for a disc around a solar 
mass star with a self-gravitating lifetime of $100,000$ years, where the blue
line adopts the assumption discussed thus far, i.e. that fragmentation occurs
only for clumps that avoid a spiral arm encounter for $t_{cool}$.  The green line 
in Figure \ref{fig:scale} is a less stringent requirement and instead involves
the assumption that the clump has been able to contract to a density
significantly larger than the mean post-shock density before it meets a spiral
arm.  At large radii, where $\beta$ is low and the spiral features are of high
amplitude this is equivalent to the criterion described above.  In the inner
disc, where the cooling time is long and the spiral features are of low
amplitude it implies a more permissive criterion for collapse; in this case we
use the arm amplitude measured in our simulations combined with the $\beta$
dependence of arm amplitude derived by \cite{Cossins1} in order to require
that the clump's density has increased by a factor of $20\sqrt{20/\beta}$
between spiral arm encounters before it is deemed to be able to collapse.

It is clear from Figure \ref{fig:scale} that the transition between fragmenting and
non-fragmenting takes place over a very narrow range of radii.  This is true
whether we require a clump to meet a density cut-off or to survive for a cooling
time before we consider it to have fragmented: indeed the two criteria are
visually nearly indistinguishable in Figure \ref{fig:scale} because by the
time one enters the high $\beta$ regime at small radius where the two criteria
diverge, the probabilities are vanishingly small.  We thus conclude that stochastic 
fragmentation cannot decrease the maximum radius at which fragmentation is 
possible by more than a few 10s of percent.

\section{Discussion}
\label{sec:discuss}

In this study, we have chosen to only perform 2D simulations where higher
resolution in $h/H$ can be obtained more easily.  It has recently
been shown that 2D simulations are problematic for studying
fragmentation, due to the effect gravitational softening has on preventing
pressure-supported collapse \citep{MYfrag}.  However, it is important to note that
the suppression of fragmentation in 2D results entirely from the modification
to the gravitational force on small scales.  Our conclusions rely entirely on
results derived from the large scale spiral structure of the disc, which is
unaffected by any small scale differences in the gravitational force law
between 2D and 3D.  As such, we expect the distribution of times between
spiral wave encounters to be unchanged in 3D.

Figures \ref{fig:tshocks} and \ref{fig:logp} show that the spiral morphology of the
disc, and hence the time between successive shock fronts, does not vary
significantly with $\beta$ or resolution.  This is consistent with the
findings of \cite{Cossins1}, who investigated the spiral geometry of 3D
gravo-turbulent accretion discs and found no $\beta$ or resolution dependence.
To the extent to which there is any trend with $\beta$ and resolution, it is
towards times between shock fronts becoming shorter at higher resolution
and/or $\beta$, which would only strength our conclusions (i.e. make
stochastic fragmentation at small radii more difficult).

Naturally, the cooling rates in a realistic disc are not likely to be well
described by cooling at constant $\beta$ once the fragments enter the regime
of non-linear collapse.  This consideration is unlikely to have a major
bearing on our analysis since what is relevant is the cooling regime while the
fragments' over-density with respect to the surroundings is still relatively
modest and where the cooling rates are therefore not expected to deviate
greatly from those appropriate to the background disc.  Following
\cite{Cossins2}, who examined the effect of temperature dependent cooling on
\emph{prompt} fragmentation, we would not expect this to facilitate
fragmentation apart from close to strongly temperature dependent features in
the opacity law, e.g. due to ice sublimation (see also \cite{Johnson2003}).  Although this
enhanced cooling of over-dense regions may bring the fragmentation boundary
inwards by a modest factor, it is unlikely to change our conclusions regarding
the \emph{sharpness} of this transition.

In focusing on the time between spiral wave passages, we have assumed that
contracting clumps are disrupted by shocks.  In apparent conflict with this
assumption, it has been shown that when fragmentation is immediate, fragments arise out of 
the dense post-shock gas that makes up the spiral arms
\citep{ArmFragCriteria}.  However, in this scenario the post-shock gas is the
seed of the instability (which then rapidly contracts to form a fragment), but
does not promote the contraction of an existing clump.  It is possible that
the dense post-shock gas is the location where the gravitational
instability is most commonly triggered.  Nonetheless, for stochastic
fragmentation to occur, this unstable patch of disc must contract
significantly from its ``birth density''.  \cite{PKstochastic} found no
evidence that the stochastic fragments formed in his simulations were driven
to fragment by spiral wave encounters.   Additionally, if it were the case that
spiral waves promoted clump contraction, it is difficult to see how any
gravitationally unstable patch of disc could fail to progress into a fragment.

Figure \ref{fig:tshocks} shows that the most common wait time between shocks
is $\sim 4-5 t_{dyn}$.  This suggests that $m\xi \approx 1$ in Equation \ref{eq:tspi},
in good agreement with the findings of \cite{Cossins1}.  Note that the
``fragmentation boundary'' measured by \cite{MeruBate1,MeruBate2,RiceCool,RiceCool2}
corresponds to the point where the probability of surviving for more than
$\beta t_{dyn}$ becomes low, not to the most common wait time.  Given this,
our results suggest the fragmentation boundary should be in the range $\beta =
7-12$, although the uncertainty in the number of over-dense regions formed per
dynamical time (see below) prevent us from placing stronger constraints.

\cite{PKstochastic} found that clumps formed in the gravo-turbulent state for
all values of $\beta$ up to $\beta \sim 50$, but was only able to find
stochastic fragmentation up to $\beta = 20$.  The probability of a simulation
fragmenting in its lifetime is approximately the probability shown in 
Figure \ref{fig:logp} (which is the probability of fragmentation 
{\it per clump}) multiplied by the number of over-dense clumps that form in 
the simulations lifetime.
\cite{PKstochastic} found roughly one clump per $\sim 50t_{dyn}$ in his $\beta=9$
simulations and found two simulations fragmented after $\sim 600$ and $\sim
800$ $t_{dyn}$ while two simulations remained stable for $1000 t_{dyn}$.  This
implies a probability of fragmentation per clump of $\sim 2/68 \approx
3\%$ ($3400/50=68$ clumps in $3400 t_{dyn}$), which is broadly consistent with 
the $\sim 10\%$ predicted by Figure \ref{fig:logp}.  By $\beta=20$, Figure 
\ref{fig:logp} predicts that only $\sim 0.5\%$ of clumps will fragment, 
suggesting fragmentation is unlikely but not impossible.

We therefore conclude that stochastic
fragmentation only modifies the ``fragmentation boundary'' in radius by $\sim
20\%$ (see Figure \ref{fig:scale}) and rules out the formation of giant planets
via any type of gravitational collapse (stochastic or otherwise), except in
the outermost parts of protoplanetary discs.  This agrees with the conclusions
of earlier disc fragmentation studies \citep{Rice05,MeruBate2,RiceCool2,PKedge}.  The  
only remaining avenue for fragmentation at low radii is via extremely large
sub Jean's length density fluctuations \citep{Hopkins}, which cannot presently
be directly probed computationally.

\section{Conclusion}

In this paper we have used simple, 2D SPH simulations of self-gravitating
discs to place hard, quantitative limits on the values of the cooling time
parameter, $\beta$, for which
massive discs can stochastically fragment.  We find that the average time any
patch of disc spends between spiral wave encounters is broadly consistent with
the findings of \cite{MYfrag} and that the stochastic nature of the disc's spiral
structure can only increase this time by at most a factor of a few.  

Consistent with earlier studies \citep{Cossins1}, we find that the
distribution of wait times between strong shocks is independent of both
resolution and $\beta$.  Furthermore, the distribution of wait times decreases
exponentially for wait times beyond $\sim 15 t_{dyn}$.  Combining this finding
with the expected $\beta \propto R^{-9/2}$ scaling of cooling time with radius, we
find that stochastic fragmentation is unable to modify the radius at which
fragmentation is possible by more than a few 10s of percent.  This finding 
restricts direct collapse as a mechanism for giant planet formation to the outer 
parts of protoplanetary discs.

\section{Materials \& Methods}
\label{sec:materials}

In the interests of reproducibility and transparency, all code and data used
in performing this work have been made freely available online at 
\url{https://bitbucket.org/constantAmateur/discfragmentation}.

\section{Acknowledgements}
\label{sec:ack}

This work benefited greatly from discussions with Giuseppe Lodato, Deborah Sijacki,
Richard Nelson, Farzana Meru, Richard Booth \& Daniel Price on issues of
both physics and numerics.  Sijme-Jan Paardekooper deserves special thanks for 
having provided advice and feedback throughout the course of this project.

Matthew Young gratefully acknowledges the support of a Poynton Cambridge
Australia Scholarship. This work has been supported·
by the DISCSIM project, grant agreement 341137 funded by the European·
Research Council under ERC-2013-ADG.

This work used the DIRAC Shared Memory Processing system at the University of
Cambridge, operated by the COSMOS Project at the Department of Applied
Mathematics and Theoretical Physics on behalf of the STFC DiRAC HPC Facility
(\url{www.dirac.ac.uk}). This equipment was funded by BIS National E-infrastructure
capital grant ST/J005673/1, STFC capital grant ST/H008586/1, and STFC DiRAC
Operations grant ST/K00333X/1. DiRAC is part of the National
E-Infrastructure.

\bibliographystyle{mn2e}
\bibliography{references}

\end{document}